\documentclass[doublecol]{epl2} 
\usepackage{amssymb}
\usepackage{amsmath}
\usepackage{bm}
\usepackage{babel}[english]
\usepackage{graphicx}
\usepackage{float}
\usepackage{MnSymbol}
\usepackage{verbatim}
\usepackage{enumerate}

\usepackage{color,soul}
\usepackage{upgreek}
\usepackage[dvipsnames,svgnames,table]{xcolor}
\usepackage{hyperref}

\hypersetup{
pdfstartview={FitH},% fits the width of the page to the window
colorlinks=true,    % false: boxed links; true: colored links
linkcolor=NavyBlue, % color of internal links
citecolor=Blue,   % color of links to bibliography
filecolor=NavyBlue, % color of file links
urlcolor=NavyBlue   % color of external links
}

\title{Multi-terminal Josephson junctions: a road to topological flux networks}
\shorttitle{Multi-terminal Josephson junctions: a road to topological flux networks}
\author{Lucila Peralta Gavensky\inst{1,2,3}
\and Gonzalo Usaj\inst{2,3} \and C. A. Balseiro\inst{2,3}}
\shortauthor{Lucila Peralta Gavensky \etal}

\institute{    
\inst{1} CENOLI, Universit\'e Libre de Bruxelles, CP 231, Campus Plaine, B-1050 Brussels, Belgium\\
  \inst{2} Centro At\'omico Bariloche and Instituto Balseiro, Comisi\'on Nacional de Energ\'ia At\'omica (CNEA)---Universidad Nacional de Cuyo (UNCUYO), (8400) Bariloche, Argentina\\
  \inst{3} Instituto de Nanociencia y Nanotecnolog\'ia (INN), Consejo Nacional de Investigaciones Científicas y T\'ecnicas (CONICET), (8400) Bariloche, Argentina
}
\abstract{Multi-terminal Josephson junctions were recently proposed as a versatile and tunable platform to emulate topological Bloch-like Hamiltonians in arbitrary dimensions.
In this perspective article, we will give a brief overview of the subject and recognize these mesoscopic devices as realizations of topological flux networks as the ones envisioned by J. E. Avron and coworkers in their seminal works on the early days of the quantum Hall effect. We  summarize the current state-of-the-art theoretical and experimental research regarding these Josephson devices, highlighting recent developments and giving an outlook on current trends.
}
\begin{document}

\maketitle
%%%%%%%%%%%%%%%%%%%%%%%%%%%%%%%%%%%%%%%%%%%%%%%%%%%%%%%%%%%%%%%%%%%%%%%%%%%%%%%%%%%%%%
\section{Introduction} 
\begin{figure}[!t]
\center
\includegraphics[width=0.9\columnwidth]{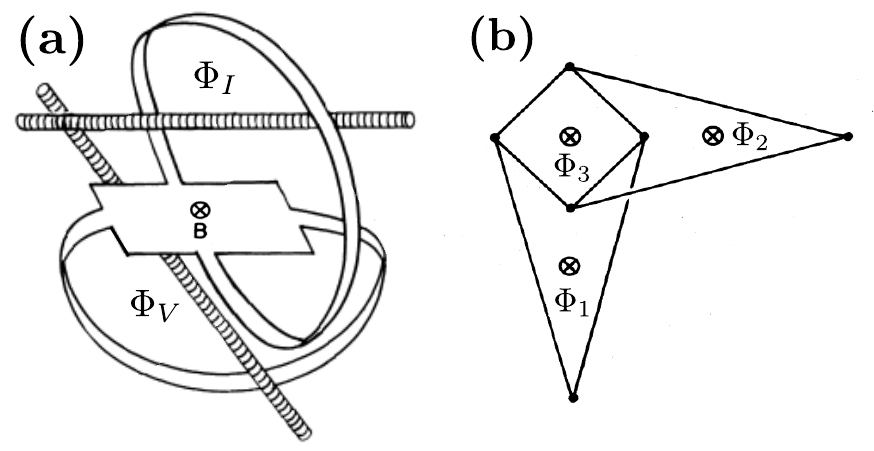}
\caption{\textbf{(a)} Hall bar with an external magnetic field $B$ in a torus geometry, adapted from Ref.~\cite{Niu1987}. The flux $\Phi_V$ is associated with an electromotive force that drives the Hall current and the flux $\Phi_I$ plays the role of an ammeter that measures the Hall current. \textbf{(b)} A three-loop network threaded by fluxes $\Phi_1,\Phi_2$ and $\Phi_3$, adapted from~\cite{Avron1988}. Its configuration mimics the one of a Hall device.}
\label{fig1}
\end{figure}

%%%%%%%%%%%%%%%%%%%%%%%%%%%%%%%%%%%%%%%%%%%%%%%%%%%%%%%%%%%%%%%%%%%%%%%%%%%%%%%%%%%%%%
Many of the foundational concepts of the field of topological condensed matter physics have been developed during the 1980's after the experimental discovery of the quantum Hall effect \cite{Klitzing1980}. The remarkably robust quantization of the Hall conductance of two-dimensional electron gases at high magnetic fields in units of $e^2/h$ was initially understood in terms of the topological properties of the one particle Bloch bands in quasimomentum space~\cite{Thouless1982}. A couple of years later, these arguments were generalized to take into account the presence of many-body interactions and disorder by considering the Hall sample to live in a torus (two loop) geometry \cite{Niu1985,Niu1987,Avron1985}, as the one sketched in Fig.~\ref{fig1}(a). The angular variables defining such a torus were identified as two Aharonov-Bohm fluxes ($\Phi_V$ and $\Phi_I$) threading the physical system and the Hall conductance as the average of the Berry curvature of the ground-state wavefunction in that parameter space~\cite{Avron2003} (also known as the many-body Chern number). 
%%%%%%%%%%%%%%%%%%%%%%%%%%%%%
%%%%%%%%%%%%%%%%%%%%%%%%%%%%%
Soon afterwards, J. E. Avron and coworkers extended the study of topologically quantized transport coefficients to the realm of mesoscopic quantum networks\cite{Avron1988,Avron1989,Avron1991}. These devices can be represented as graphs [see for example Fig.\ref{fig1}(b)] with $D$ loops threaded by independent, externally controllable, flux tubes $\Phi_l$ ($l=1,\hdots,D$). The synthetic parameter space defined by these fluxes can be identified with a D-dimensional torus $\{\Phi_l\} \in \mathbb{T}^{D}$ with a fundamental period given by the flux unit $\Phi_0=hc/q$, where $q$ is the charge of the carriers in the device. These carriers are constrained to move along the links of the network, and their dynamics is governed by an appropriate Schr\"odinger equation defined in the corresponding multiply connected domain. In Ref.~\cite{Avron1988} concrete Hamiltonian models describing such networks were analyzed, either by considering the vertices of each graph as molecular sites with a tight-binding like description or by analyzing the dynamics of free particles moving along one-dimensional wires in the system~\cite{Avron1991}.

Considering a given Hamiltonian $\hat{H}$, the operator associated to the charge current flowing through loop $k$ is given by $\hat{I}_k = -c\frac{\partial\hat{H}}{\partial\Phi_k}$. With the aim of probing the transport properties of the network, some of its constituent fluxes can be externally driven in time.
When doing so in an adiabatic manner, the expectation value of the $k$-th current can be obtained up to first derivatives of the perturbation as
%%%%%%%%%%%%%%%%%%%%%%%%
\begin{eqnarray}
\langle \hat{I}_k (t)\rangle &=& -c \frac{\partial E_0[\bm{\Phi}(t)]}{\partial \Phi_k} + \frac{hc}{2\pi}\sum_l \mathcal{F}_{kl}[\bm{\Phi}(t)]\,\dot{\Phi}_l,
\label{eqI}
\end{eqnarray}
%%%%%%%%%%%%%%%%%%%%%%%%
where $E_0[\bm{\Phi}(t)]$ is the instantaneous ground state energy of the system and
%%%%%%%%%%%%%%%%%%%%%%%%
\begin{equation}
 \mathcal{F}_{kl}[\bm{\Phi}(t)] = i\left(\langle \partial_{k} \psi_0[\bm{\Phi}(t)]|\partial_{l} \psi_0[\bm{\Phi}(t)]\rangle - k \leftrightarrow l\right)\,,
\end{equation}
%%%%%%%%%%%%%%%%%%%%%%%%
is the Berry curvature of the corresponding eigenstate $|\psi_0\rangle$ evaluated at $\bm{\Phi}(t)=(\Phi_1(t),\hdots,\Phi_D(t))$---assumed to be non-degenerate. We have used here the convention $\partial_{\Phi_k} \equiv \partial_k$. The above derivation requires the adiabatic postulate to hold, which can be insured for sufficiently slow driving protocols and as long as there is a gap between the ground state of the system and the excitation spectrum. As pointed out in Ref.~\cite{Avron1988}, this transport analysis strongly relies on quantum coherence of the wave function over the entire multiply connected device. Such phase rigidity is expected to be present in superconductors over macroscopic lengths and in normal metals under more stringent conditions on length and temperature scales. 

The first term in Eq.~\eqref{eqI} can be identified as a persistent current, which may be present even in the absence of the driving. If the fluxes vary linearly in time, a constant electromotive force (emf) $\mathcal{E}_l = -\frac{1}{c}\dot{\Phi}_l$ is generated around each loop, allowing for the definition of a conductance matrix with elements given by $G_{k l}[\bm{\Phi}(t)] = \frac{\partial\langle \hat{I}_k (t)\rangle }{\partial \mathcal{E}_l}$. These transport coefficients relate the current flowing around loop $k$ to the emf around loop $l$. Remarkably, when averaging them over a $\Phi_0 \times \Phi_0$ square slice in flux space, they become quantized in units of $q^2/h$
%%%%%%%%%%%%%%%%%%%%%%%%
\begin{eqnarray}
\overline{G}_{k l} = \frac{1}{\Phi_0^2} \int_0^{\Phi_0}d\Phi_k\int_0^{\Phi_0}d\Phi_l\,\frac{\partial\langle \hat{I}_k (t)\rangle }{\partial \mathcal{E}_l} = -\frac{q^2}{h} C_{kl}\,,
\label{GAvron}
\end{eqnarray}
%%%%%%%%%%%%%%%%%%%%%%%
with $C_{kl}$ the Chern number of the ground state. We have here used the Chern theorem in order to relate the integral of the Berry curvature over a closed periodic manifold with an integer topological index
%%%%%%%%%%%%%%%%%%%%%%%
\begin{equation}
C_{kl} = \frac{1}{2\pi} \int_0^{\Phi_0}d\Phi_k\int_0^{\Phi_0}d\Phi_l\, \mathcal{F}_{kl}[\bm{\Phi}]\,,
\end{equation}
%%%%%%%%%%%%%%%%%%%%%%%
and that the persistent current term averages to zero. The value of this topological number depends on the remaining (unaveraged) fluxes with a $\Phi_0$-periodicity. Interestingly, these integers do not necessarily classify a property of a given physical sample, as in the quantum Hall setup of Fig.~\ref{fig1}(a), but rather the topology of the whole network itself~\cite{Avron1988}: the leads connecting each vertex, the number of holes and their geometrical layout. Non trivial averaged conductances generically arise in networks with at least three independent fluxes $D\geq 3$~\cite{Avron1988}, a dimensionality constraint mathematically rooted in the co-dimension theorem of Von Neumann and Wigner~\cite{Wigner1929}. Indeed, three independent variables are needed to fine tune degeneracies in the energy spectrum, which are known to act as sources or sinks of Berry curvature. In the particular case of $D=3$, the degeneracies occur at points in the parameter space, which are generically dubbed Weyl singularities. The Hamiltonian near a two-band touching point $\bm{\Phi}_W$ can be approximated with a two-by-two matrix as
%%%%%%%%%%%%%%%%%%%%%%%%%%%
\begin{equation}
\label{weyl}
\hat{H}_{W}= \bm{h}(\bm{\Phi}_W+\delta\bm{\Phi})
\cdot\bm{\sigma}=\sum_{k,i}\delta \Phi_{k} \mathbb{M}_{ki}\sigma_i,    
\end{equation}
%%%%%%%%%%%%%%%%%%%%%%%%%%%
where $\mathbb{M}_{ki} = \partial_{\Phi_k}h_i\rvert_{\bm{\Phi}_W}$ and $\sigma_i$ are the Pauli matrices ($i=x,y,z$). We have here neglected the terms proportional to the identity matrix, since they have no influence on the existence of the degeneracy nor on its topological classification. Weyl points can be considered as monopoles of topological charge which come in two different flavours or chiralities, determined by $\chi = \textrm{sign}(\textrm{det}\mathbb{M})$.
%%%%%%%%%%%%%%%%%%%%%%%%%%%%%%%%%%%%%%%%%%%%%%%%%%%%%%%%%%%%%%%%%%%%%%%%%%%%%%%%%%%%%%%
\section{Multi-terminal Josephson junctions}
%%%%%%%%%%%%%%%%%%%%%%%%%%%%%%%%%%%%%%%%%%%%%%%%%%%%%%%%%%%%%%%%%%%%%%%%%%%%%%%%%%%%%%
%%%%%%%%%%%%%%%%%%%%%%%%%%%%%%%%%%%%%%%%%%%
\begin{figure}[t]
\center
\includegraphics[width=0.95\columnwidth]{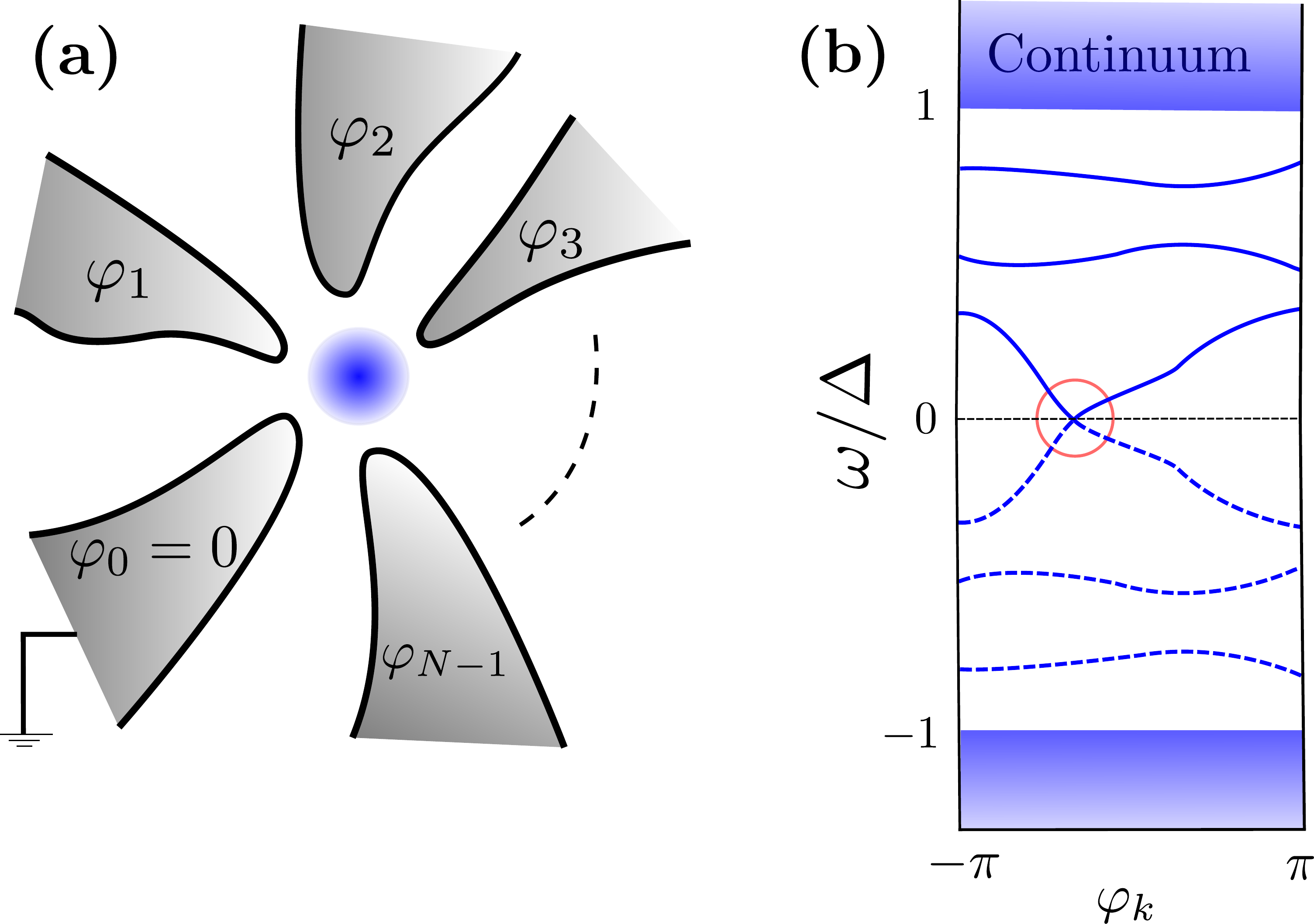}
\caption{\textbf{(a)} Sketch of a multi-terminal Josephson junction, where $N$ superconducting leads with phases $\varphi_l$ ($l=0,1,\hdots,N-1$) are coupled through a central scattering region. \textbf{(b)} Illustration of a generic Bogoliubov-de Gennes quasiparticle energy spectrum of these devices as a function of one of the phase $\varphi_k$. We indicate with a red circle a zero energy crossing.}
\label{fig2}
\end{figure}
%%%%%%%%%%%%%%%%%%%%%%%%%%%%%%%%%%%%%%%%%%
A few years ago, these ideas~\cite{Avron1988,Avron1989} experienced a renaissance when multi-terminal Josephson junctions (JJs) were proposed as a novel platform for topological engineering~\cite{Riwar2016}. A typical device is sketched in Fig.~\ref{fig2}(a), where $N$ s-wave superconducting leads (grey) are coupled through a central scattering region (blue). 
Each of these terminals is characterized by a complex mean-field order parameter $\Delta_l e^{i\varphi_l}$ and gauge invariance allows to choose a gauge in which $\varphi_0=0$, leaving $N-1$ free phase parameters $\bm{\varphi}=(\varphi_1,\varphi_2,\hdots,\varphi_{N-1})$ in the system. For simplicity, we will take $\Delta_l = \Delta$.
In these Josephson devices, the phase differences are usually controlled with superconducting loops threaded by external magnetic fluxes $\Phi_l$ so that $\varphi_l = 2\pi \Phi_l / \Phi_0^s$ with $\Phi_0^s = hc/2e$ the superconducting flux quantum. One can accordingly identify these junctions as (N-1)-dimensional flux networks where the effective charge $q=2e$ is the Cooper pair charge. The quasiparticle energy spectrum obtained when diagonalizing the Bogoliubov-de Gennes (BdG) Hamiltonian describing these junctions usually consists on a continuum spectrum above the superconducting gap $\Delta$ and a set of discrete levels (from now on indexed by the label $\alpha$) living inside the gap, as illustrated in Fig.~\ref{fig2}(b). The microscopic origin of the discreteness of this last part of the spectrum is rooted in the presence of Andreev reflection processes at the superconducting interfaces: incident electrons with energies below the superconducting gap are reflected as holes and viceversa, leading to a confined motion in the central scattering region of the nanostructure which ge\-ne\-ra\-tes the so-called Andreev bound states (ABS)~\cite{Andreev1964,Klapwijk2004,Nazarov2009}. These Bogoliubov quasiparticles bear a certain dispersion relation $\varepsilon_{\alpha}(\bm{\varphi})$ as a function of the phase differences of the superconducting leads, with a functional form that depends both on the properties of the terminals and the scattering region that connects them~\cite{Beenakker1991}. The phase differences of the leads define a closed ($2\pi$-periodic) manifold and hence play the role of an effective Brillouin zone (BZ) in the problem. Indeed, a formal analogy between the ABS of $N$-terminal JJs and the Bloch band structure of materials in $N-1$ dimensions can be established, with the superconducting phases $\varphi_l$ playing the role of the quasimomenta components $k_l$ in a crystalline structure. In this way, the number of synthetic dimensions of the problem can be simply increased by adding more superconducting leads. 

The key idea of Ref.~\cite{Riwar2016} is that the ABS spectrum may host Weyl singularities [as the one circled in red in Fig.~\ref{fig2}(b)] if there are at least three independent superconducting phases ($N\geq 4$), as expected from Wigner's co-dimension theorem. Note that a zero energy crossing of the eigenstates of the BdG Hamiltonian means that the excitation spectrum of the junction is gapless: there is a degeneracy between the ground state wavefunction of the junction and a state with one extra/less quasiparticle. 
%%%%%%%%%%%%%%%%%%%%%%%%%%%%%%%%%%%%%%%%%%%%%%%%%%%%%%%%%%%%%%
\begin{figure}[t]
\center
\includegraphics[width=0.95\columnwidth]{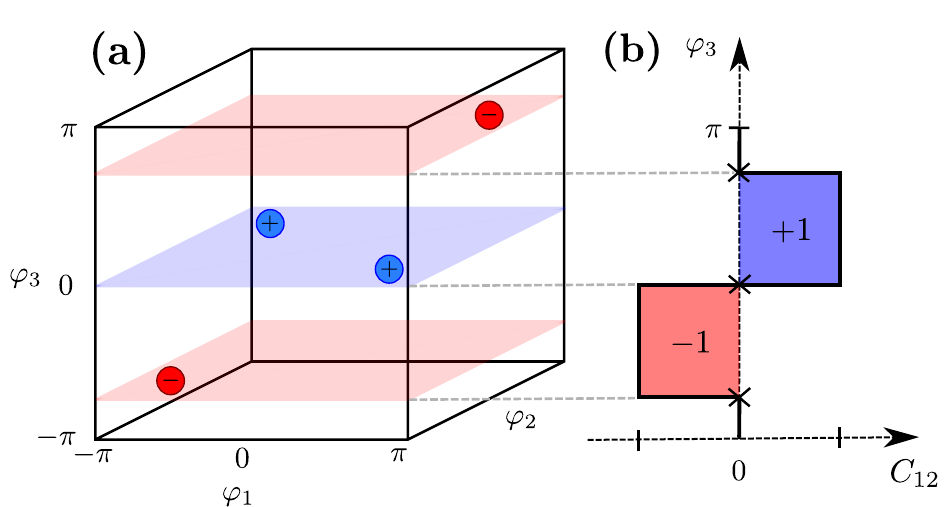}
\caption{\textbf{(a)} Sketch of the position of four Weyl points in the effective three-dimensional Brillouin zone of a generic four-terminal Josephson junction. We mark with blue Weyl points with positive topological charge ($\chi = +1$) and with red the ones with negative topological charge ($\chi = -1$). \textbf{(b)} Chern number $C_{12}$ defined in the two-dimensional moving plane $(\varphi_1,\varphi_2)$ as a function of the phase difference $\varphi_3$. }
\label{W4JJ}
\end{figure}
%%%%%%%%%%%%%%%%%%%%%%%%%%%%%%%%%%%%%%%%%%%%%%%%%%%%%%%%%%%%%%

In the case of 4-terminal junctions, the synthetic BZ is three-dimensional and the Weyl singularities must be points in this parameter space. Due to the Nielsen-Ninomiya theorem~\cite{Nielsen1981}, the sum of the charges of Weyl monopoles in a 3D BZ must vanish, so Weyl points can never appear in isolation: they must come in pairs with opposite charges. Furthermore, time reversal invariance in the scattering region implies that $\varepsilon_{\alpha}(\bm{\varphi})=\varepsilon_{\alpha}(-\bm{\varphi})$, so if there is a Weyl crossing at $\bm{\varphi}_{W}$ there must be another one at $-\bm{\varphi}_{W}$. Even more, since the chirality $\chi$ of a Weyl point is  even under time reversal, both must have the same topological charge. This guarantees that Weyl points must come in groups of four, as sketched in Fig.~\ref{W4JJ}\textbf{(a)}. By randomly generating scattering matrices from the circular orthogonal ensemble to describe such junctions, the authors of Ref.~\cite{Riwar2016} indeed showed that a significant fraction of them yielded zero-energy Weyl points in the BdG spectrum, with the number of singularities increasing when increasing the number of channels in each terminal. A topological classification can be made by taking, with\-out loss of generality, the phase difference $\varphi_3$ as a control parameter.  In this way, it is possible to define a Chern number $C_{12}(\varphi_3)$ in the two-dimensional submanifold defined by the moving plane with coordinates $(\varphi_1,\varphi_2)$. By denoting the wavefunctions of the Andreev bound states as $|\phi_{\alpha}(\bm{\varphi})\rangle$, this topological invariant takes the form
%%%%%%%%%%%%%%%%%%%%%%%%%%%
\begin{equation}
C_{12}(\varphi_3)=\sum_{\alpha\,|\,\varepsilon_{\alpha} <0} C_{12}^{\alpha}(\varphi_3),
\end{equation}
%%%%%%%%%%%%%%%%%%%%%%%%%%
with
%%%%%%%%%%%%%%%%%%%%%%%%%%
\begin{equation}
 C_{12}^{\alpha}(\varphi_3)= \frac{i}{2\pi}\!\int_{-\pi}^{\pi}\!\!d\varphi_1\int_{-\pi}^{\pi}\!\!d\varphi_2 \left(\langle \partial_{\varphi_1}\phi_{\alpha}|\partial_{\varphi_2}\phi_{\alpha}\rangle\!-\!\varphi_1\leftrightarrow\varphi_2\right).
\end{equation}
%%%%%%%%%%%%%%%%%%%%%%%%%
The sum of the Chern numbers $C_{12}^{\alpha}$  of the $\alpha$-th ABS of the BdG Hamiltonian up to the Fermi level must be interpreted as the Chern number of a state with no quasiparticles at zero temperature. We plot in Fig.~\ref{W4JJ}(b) how this topological index would look like as a function of $\varphi_3$ for a configuration of Weyl points as the one sketched in panel (a).

A natural follow up question is whether this topological classification has any observable consequences in the transport properties of these devices. Following the same line of reasoning as in the previous section, one may think of driving one of the phases $\varphi_l$ in time and studying the corresponding adiabatic response in a different terminal. In these superconducting networks, this may be done by applying a bias voltage $V_{l}$ in terminal $l$ so that the corresponding superconducting phase acquires a time dependence in accordance with the Josephson relation $\dot{\varphi}_{l}(t)=2 e V_{l}/\hbar$. If the ABS spectrum is gapped and the bias voltage is small compared to this gap, with the same logic as the one used in Eq.~\eqref{eqI}, one can show (see Ref.~\cite{Riwar2016}) that the ABS contribution to the current in terminal $k$ at zero temperature is given by
%%%%%%%%%%%%%%%%%%%%%%
\begin{equation}
\langle I_{k}(t)\rangle = \frac{2 e}{\hbar}\sum_{\alpha\,|\,\varepsilon_{\alpha} <0} \frac{\partial \varepsilon_{\alpha}[\bm{\varphi}(t)]}{\partial \varphi_{k}} - 2e \sum_{l,\alpha\,|\,\varepsilon_\alpha <0}\mathcal{F}^{\alpha}_{kl}[\bm{\varphi}(t)]\,\dot{\varphi}_{l}(t)\,,   
\label{ad_Nazarov}
\end{equation}
%%%%%%%%%%%%%%%%%%%%%%%
with $\mathcal{F}^{\alpha}_{kl}$ the Berry curvature of the $\alpha$-th bound state evaluated at the instantaneous phase differences $\bm{\varphi}(t)$. In this expression, spin degeneracy is assumed and a summation over spin indexes has already been made. By defining a transconductance matrix with elements $G_{kl} = \frac{\partial \langle I_k (t) \rangle}{\partial V_l}$ and phase averaging this expression one readily finds that
%%%%%%%%%%%%%%%%%%%%%%%
\begin{equation}
\overline{G}_{kl} = -\frac{4e^2}{h}\sum_{\alpha\,|\,\varepsilon_{\alpha} <0} C^{\alpha}_{kl}, 
\label{CNazarov}
\end{equation}
%%%%%%%%%%%%%%%%%%%%%%%
with $C^{\alpha}_{kl}$ the Chern number of the $\alpha$-th ABS in the sub-manifold defined by the phases $(\varphi_k,\varphi_l)$. Note that Eq.~\eqref{CNazarov} is in complete analogy with Eq.~\eqref{GAvron}. The above derivation can also be generalized to the case where spin-rotation symmetry is broken due to, for example, Zeeman or spin-orbit coupling effects. In such circumstances, zero energy Weyl points split and move to finite energy while preserving their topological charge~\cite{Yokoyama2015}. Although the gap in the spectrum is closed along a contour encircling these singularities, it was argued that the phase averaged trans\-conductance could remain quantized provided fermion parity is preserved in the junction~\cite{Yokoyama2015,Riwar2016}. 

The existence of a phase averaged topologically quantized transconductance has been subsequently tested in Ref.~\cite{Eriksson2017} by performing transport simulations in a voltage biased four terminal junction. These findings were followed up by numerous theoretical works that have thoroughly analyzed the emergence and evolution of Weyl nodes in the ABS spectra of a variety of JJs~\cite{Yokoyama2017,Xie2018,Klees2021}. It was soon realized that three terminal Josephson devices (formally equivalent to two-dimensional bulk materials) could also host topological Andreev states when time-reversal symmetry is broken in the central scattering region~\cite{Meyer2017,Xie2017}. This may be achieved by piercing the junction area with an additional magnetic flux or by driving the system with a periodic time-dependent microwave field~\cite{PeraltaGavensky2018}. In this last case, the well-known tools of Floquet engineering~\cite{Oka2019} were introduced to generate topologically non-trivial Andreev states by photon dressing. Additionally, several works have also generalized the problem to heterostructures built with superconducting reservoirs in the presence of $p$-wave correlations, where Majorana zero modes could ultimately emerge and get coupled to each other~\cite{Houzet2019,PeraltaGavensky2019,Xie2019,Kotetes2019,Sakurai2020,Meyer2021}. In this regime, single electrons can be transferred between terminals, resulting in a phase averaged transconductance quantized in units of $2e^2/h$ instead of $4e^2/h$.
%%%%%%%%%%%%%%%%%%%%%%%%%%%%%%%%%%%%%%%%%%%%%%%%%%%%%%%%%%%%%
\section{Locally accessing quantum geometry in flux-space}
%%%%%%%%%%%%%%%%%%%%%%%%%%%%%%%%%%%%%%%%%%%%%%%%%%%%%%%%%%%
An interesting advantage of these devices, compared to other solid-state platforms, is that it is conceptually much easier to probe local geometrical properties of quantum states in the artificial Brillouin zone spanned by the phase-differences. In principle, it is possible to perform measurements at each effective quasimomenta by simply manipulating the external fluxes of the junction in an independent manner. The full quantum geometry of the Andreev bound states in this synthetic flux space is encoded in the quantum geometric tensor, defined as
%%%%%%%%%%%%%%%%%%%%%%%%
\begin{equation}
\mathcal{Q}^{\alpha}_{kl}[\bm{\varphi}]=\langle \partial_{\varphi_k}\phi_{\alpha} |\left(1 - |\phi_{\alpha}\rangle\langle \phi_{\alpha}|\right)|\partial_{\varphi_l} \phi_\alpha\rangle.
\label{qgt}
\end{equation}
%%%%%%%%%%%%%%%%%%%%%%%
On the one hand, the real part of this tensor provides for the Fubini-Study metric $g^{\alpha}_{kl}[\bm{\varphi}] = \textrm{Re}(\mathcal{Q}^{\alpha}_{kl}[\bm{\varphi}])$, which contains information on the ``distance'' between two quantum states. On the other hand, the imaginary part gives direct access to the Berry curvature $\mathcal{F}^{\alpha}_{kl}[\bm{\varphi}] = -2\textrm{Im}(\mathcal{Q}^{\alpha}_{kl}[\bm{\varphi}])$, which provides for a measure of quantum geometrical phases acquired along closed paths. In Ref.~\cite{Ozawa2018}, a scheme to measure the components of this tensor in a generic parameter space was developed. It essentially consists on modulating the corresponding parameters periodically in time and monitoring the excitation rates between the ground and excited states. This same strategy was adapted to the case of multi-terminal Josephson devices~\cite{Klees2020,Klees2021}, where it was proposed that microwave spectroscopy could reveal all the information contained in the geometrical object of Eq.~\eqref{qgt}. In these nanostructures, the protocol involves applying ac-drives in the external fluxes, so as to be able to modulate in time the phase differences between different reservoirs. We note that in Ref.~\cite{PeraltaGavensky2018}, an alternative scheme to locally probe the Berry curvature by performing synchronic transport measurements was also discussed.
%%%%%%%%%%%%%%%%%%%%%%%%%%%%%%%%%%%%%%%%%%%%%%%%%%%%%%%%%%%%%%%%%%%%%%%%%%%%%%%%%%%%%%
\section{On the effect of the quasiparticle continuum}
\begin{figure}[t]
\center
\includegraphics[width=0.95\columnwidth]{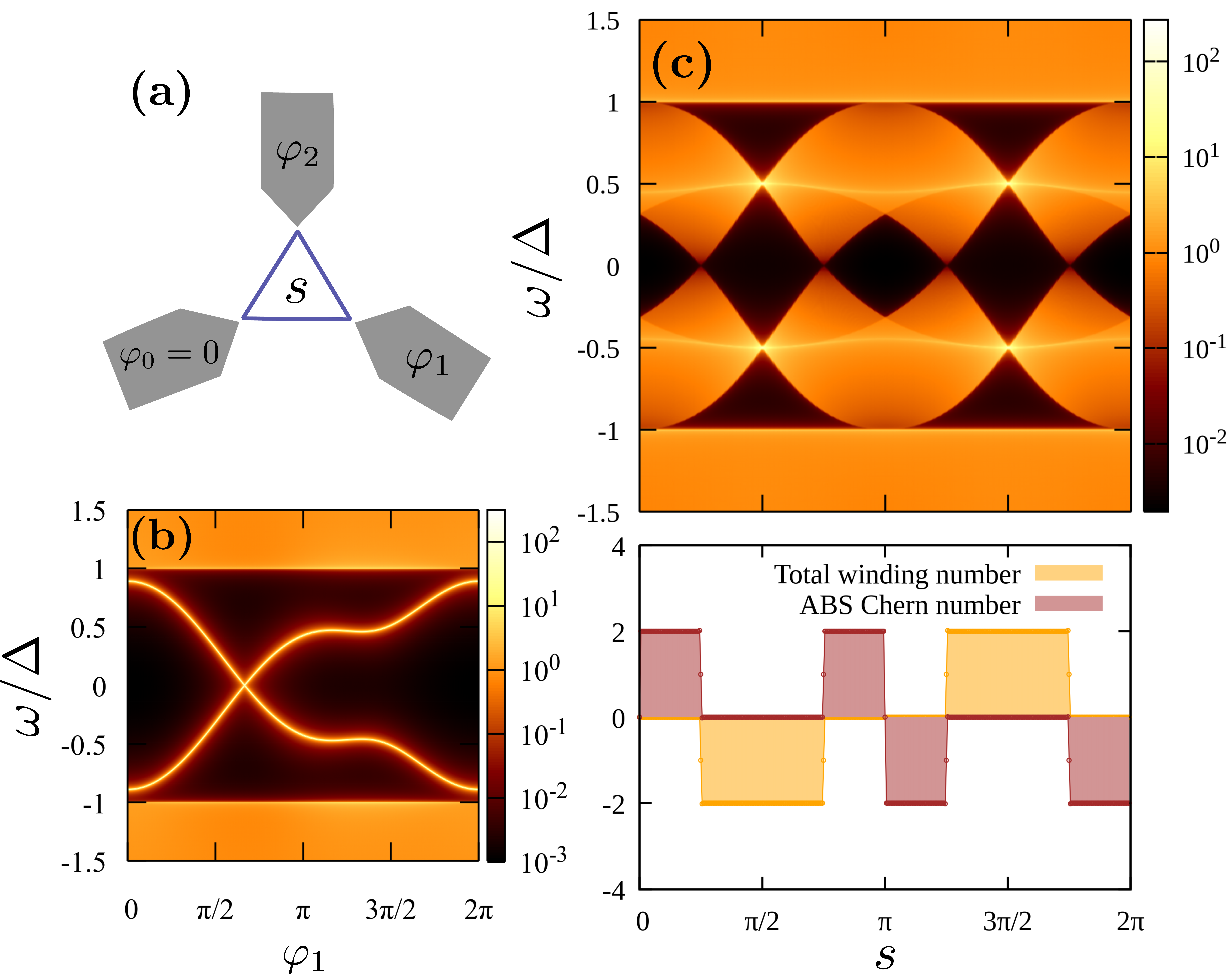}
\caption{\textbf{(a)} Schematic setup of a three-terminal Josephson junction pierced with a central magnetic flux $\Phi_s = \Phi_0 s/2\pi$ with $\Phi_0=hc/e$. \textbf{(b)} Spectral density  of the junction [$(-1/\pi) \mathrm{Im}(\mathrm{Tr}\mathcal{G}_J(\omega,\varphi_1))$] along the path $\varphi_1 = -\varphi_2$ for $s=0.78$. \textbf{(c)} Top: Phase integrated spectral density of the junction as a function of $s$. Bottom: ABS Chern number and total winding number as a function of $s$. The difference between them corresponds to the continuum contribution.}
\label{FM}
\end{figure}

%%%%%%%%%%%%%%%%%%%%%%%%%%%%%%%%%%%%%%%%%%%%%%%%%%%%%%%%%%%%%%%%%%%%%%%%%%%%%%%%%%%%%%
The analogy between the Bloch band structure and the quasiparticle energy spectrum of multi-terminal JJs, although appealing, is incomplete. Indeed, the continuum spectrum above the superconducting gap [see Fig.~\ref{fig2}(b)] also participates in the transport properties of these devices. In this sense, Eq.~\eqref{ad_Nazarov} is not entirely accurate, since it disregards the contribution of these states to the current. This problem was highlighted in Ref.~\cite{Meyer2017}, where the authors studied a setup as the one sketched in Fig.~\ref{FM}(a): three superconducting reservoirs coupled via a normal region threaded by an external magnetic flux. In their particular setting, the numerically calculated phase averaged transconductance did not match the Chern number of the ABS, a fact which was attributed to a non-trivial contribution to the quantized transconductance coming from the quasiparticle conti\-nuum.
%%%%%%%%%%%%%%%%%%%%%%%%%%%%%%%%%%%%%%%%%%%%%%%%%%%%%%%%%%%%%% 
%%%%%%%%%%%%%%%%%%%%%%%%%%%%%%%%%%%%%%%%%%%%%%%%%
In order to properly account for the continuum contribution, the adiabatic expansion of the Josephson current needs to be formulated on more general grounds. The geometrical contribution, identified as the Berry curvature of the ground state, is more conveniently written in terms of the BdG Green's functions~\cite{PeraltaGavensky2018,PeraltaGavensky2019,PeraltaGavensky2022}.
Namely,
%%%%%%%%%%%%%%%%%%%%%%%%%%%%%%%%%%%%%
\begin{eqnarray}
\langle I^{\mathrm{geo}}_{k}(t)\rangle\!\!&=&\!\!\!\!\!-e\sum_{\nu'}\int_{-\infty}^{\infty}\frac{d\omega}{4\pi}\times\\
\nonumber
&&\!\!\!\!\!
\mathrm{Tr}\Big[\widetilde{\mathcal{G}}^{-1}\partial_{\varphi_{k}}\widetilde{\mathcal{G}}^{}\cdot\widetilde{\mathcal{G}}^{-1}\partial_{\varphi_{l}}\widetilde{\mathcal{G}}^{}\cdot\widetilde{\mathcal{G}}^{-1}\partial_{\omega}\widetilde{\mathcal{G}}^{}-k \leftrightarrow l\Big]\,\dot{\varphi}_{l}\,.
\end{eqnarray}
%%%%%%%%%%%%%%%%%%%%%%%%%%%%%%%%%%%
Here, $\widetilde{\mathcal{G}}^{}(\omega,t)=\widetilde{\mathcal{G}}^{}(\omega,\{\varphi_l(t)\})$ is the fully adiabatic causal Green function obtained from the instantaneous BdG Hamiltonian~\cite{PeraltaGavensky2019}.
After averaging over the Brillouin zone as discussed before, we obtain that
%%%%%%%%%%%%%%%%%%%%%%%%%%%%%%%%%
%\begin{equation}
$\overline{G}_{kl}=-(2e^2/h)\sum_{l}\mathcal{N}_{kl}$
%\end{equation}
%%%%%%%%%%%%%%%%%%%%%%%%%%%%%%%%%%
with
%%%%%%%%%%%%%%%%%%%%%%%%%%%%%%%%%
\begin{eqnarray}
\label{Nkl}
\mathcal{N}_{kl}\!\!\!&=&\!\!\!\frac{1}{8\pi^2}\int_{-\pi}^{\pi}d\varphi_{k}\int_{-\pi}^{\pi}d\varphi_{l}\int_{-\infty}^{\infty}d\omega \times\\
&&
\nonumber
\mathrm{Tr}\Big[\widetilde{\mathcal{G}}^{-1}\partial_{\varphi_{\nu}}\widetilde{\mathcal{G}}^{}\cdot\widetilde{\mathcal{G}}^{-1}\partial_{\varphi_{\nu'}}\widetilde{\mathcal{G}}^{}\cdot\widetilde{\mathcal{G}}^{-1}\partial_{\omega}\widetilde{\mathcal{G}}^{}-k \leftrightarrow l\Big]\,,
\end{eqnarray}
%%%%%%%%%%%%%%%%%%%%%%%%%
the winding number of the Green's function $\widetilde{\mathcal{G}}$.  
If the system is gapped, this quantity is known to be an integer number~\cite{Volovik2009}. 
$\mathcal{N}_{kl}$ includes the contribution of both the ABS and the continuum spectrum on equal footing. In fact, the two contributions cannot be separated in general, except when the ABS are isolated from the continuum. 
Moreover, since the trace involves a sum over spin variables, whenever there is an equal spin contribution we expect the transconductance to be quantized in units of $4e^2/h$. An alternative expression was also derived in terms of the scattering matrix in Ref. \cite{Repin2019}.

For calculation purposes, it is convenient to gauge the superconducting phases to the tunneling matrix elements between the leads and the normal region.
%or, in the example of Fig. \ref{FM}($a$), simply between the leads
In this way, the phase dependence appears only on the tunneling Hamiltonian and the expression for the topological invariant can be written~\cite{PeraltaGavensky2019} in terms of the junction’s Green function  $\mathcal{G}_J(\omega, \{\varphi_l\})$ which involves only the region of interest. 
To illustrate the procedure, we consider three $s$-wave superconducting reservoirs (treated in the wide band limit) which are coupled to each other by (single site) normal hopping terms enclosing a magnetic flux $\Phi_s$ as depicted in  Fig. \ref{FM}(a). We assume spin degeneracy and work on the Nambu basis, so that $\mathcal{G}_J(\omega, \{\varphi_l\})$ is a $6\times6$ matrix corresponding to the three coupled ending sites of the superconducting leads. The spectral density, $(-1/\pi) \mathrm{Im}[\mathrm{Tr}\mathcal{G}_J(\omega,\{\varphi\})]$,  as a function of $\varphi_1$ ($=-\varphi_2$) for a particular value of $\Phi_s$ is shown in Fig.~\ref{FM}(b), a situation where  a Weyl singularity is present on the ABS spectrum. The phase integrated spectral density as a function of $s=2 \pi \Phi_s / \Phi_{0}$, and the corresponding ABS Chern number and total win\-ding number (Eq.~\eqref{Nkl}), are presented in Fig.~\ref{FM}(c). It is important to notice here that the total winding number only changes when there is a gap closing at the Fermi energy ($\omega=0$) while the ABS Chern number can also do it when the ABS touches the continuum ($s=0,\pi$). It is only the former quantity that is related to the transconductance---the result in Fig.~\ref{FM}(c) agrees well with the numerically calculated transconductance~\cite{Meyer2017}.
%%%%%%%%%%%%%%%%%%%%%%%%%%%%%%%%%%%%%
\section{Topological Hamiltonian}
%%%%%%%%%%%%%%%%%%%%%%%%%%%%%%%%%%
There is another way to calculate the topological invariant that, in some cases,  allows to separate the contribution of the ABS and the quasiparticle continuum. As it was recently demonstrated \cite{Wang2012,Wang2013}, under quite general assumptions, all the topological information of the Green's function $\widetilde{\mathcal{G}}(\omega)$ is encoded in the so-called `topological Hamiltonian' defined as $H_{\mathrm{top}}\equiv -\widetilde{\mathcal{G}}^{-1} (\omega = 0)$.   
The key idea is that if the system is gapped and $\widetilde{\mathcal{G}}(\omega)$ well behaved, the latter can be smoothly deformed to $\mathcal{G}_{\mathrm{eff}}(\omega) = (\omega+\widetilde{\mathcal{G}}^{-1} (\omega = 0))^{-1}$ preserving its topology. 
In that case, the winding number of the Green’s function can be calculated from the Chern numbers associated to each of the eigenstates of $H_{\mathrm{top}}$. 

In the case of JJs, the eigenstates of $H_{\mathrm{top}}^J= -\mathcal{G}_J^{-1} (\omega = 0)$ should then reflect the topological nature of both the ABS and the continuum spectrum. In the example of Fig.~\ref{FM},  $H_{\mathrm{top}}^J$ has six particle-hole symmetric eigenstates but only those below the Fermi energy need to be considered.
The Chern number associated to the ABS is obtained from the eigenstate whose energy is closer to the Fermi energy (Fig.\ref{FM}(c)), while the one related to the continuum is just the sum of the Cherns of the two remaining modes. The sum of all these contributions is precisely the winding number of the full frequency dependent Green’s function obtained before, revealing the validity of the method~\cite{PeraltaGavensky2022}. 
Notably, even though the eigenenergies obtained from $H_{\mathrm{top}}^J$ do not represent the physical spectrum, they constitute a reduced Hilbert space that accurately describes the closing of the gaps and the topology of the original problem. 
Therefore, when applicable, this method provides a much simpler way of obtaining the topological invariants without the need of carrying out an integral over frequency as in Eq.~\eqref{Nkl}. 
%%%%%%%%%%%%%%%%%%%%%%%%%%%%%%%%%%%%%%%%%%%%%%
\section{Experimental advances} 
\begin{figure}[t]
\center
\includegraphics[width=\columnwidth]{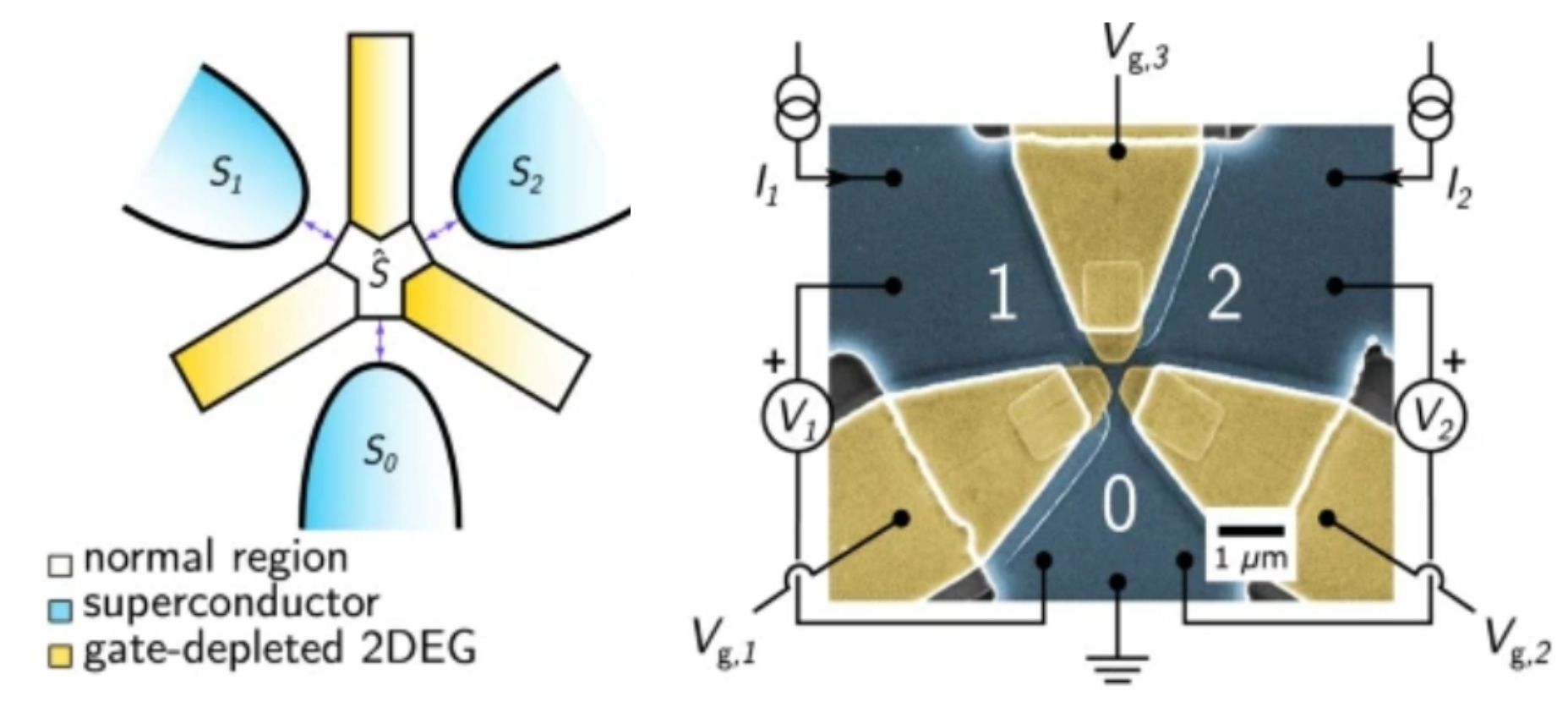}
\caption{Left Panel: Schematic of a Y-shaped three-terminal  Josephson junction in a split-gate quantum point contact geometry. Right panel: False-color SEM image of the actual device with individually tunable quantum point contacts showing measurement schematic.
Extracted from Ref.~\cite{Graziano2022}. }
\label{exp}
\end{figure}

%%%%%%%%%%%%%%%%%%%%%%%%%%%%%%%%%%%%%%
Weyl singularities in the ABS spectrum of multi-terminal Josephson devices have not been experimentally observed so far nor phase-averaged quantized transconductances measured. Contrary to the well established two-terminal JJs case, the design, fabrication and ultimate control of multi-terminal junctions is much less explored, with several experimental challenges ahead. 
Substantial progress has been recently made in this direction, partially fuelled by the perspective of engineering synthetic topology.
Several groups have already been able to develop mesoscopic JJs with three or four terminals that exhibited superconducting phase coherent transport in all-metal junctions~\cite{Pfeffer2014}, hybrid semiconductor-superconductor heterostructures~\cite{Cohen2018,Graziano2020,Pankratova2020,Graziano2022} or proximitized graphene\cite{Draelos2019,Arnault2021}. Even a
phase-tu\-na\-ble three-terminal Josephson interferometer has been realized~\cite{Strambini2016,Vischi2017}. 
One of the challenges is to have the ability to control 
the number of conductance channels between the superconducting terminals, while keeping them highly transmissive---they need to be just a few in order to have enough experimental resolution to access the ABS spectrum. A step forward in this direction was recently done in Ref.~\cite{Graziano2022} where the authors were able to control each channel of a three-terminal junction down to a single mode using a split-gate quantum-point contact geometry, see Fig.~\ref{exp}.
Other aspects that need further investigation are related to the capability of independently controlling the superconducting phases, that is magnetic fluxes, while avoiding mutual inductances and also reducing  quasi-particle poisoning~\cite{Zgirski2011} and the presence of thermal excitations. The latter is important to guarantee the ground state occupation and hence the quantization of the trans\-con\-duc\-tan\-ce. 
%%%%%%%%%%%%%%%%%%%%%%%%%%%%%%%%%%%%%%%%%%%%%%%%%%%%%%%%%%%%%% 
%%%%%%%%%%%%%%%%%%%%%%%%%%%%%%%%%%%%%%%%%%%%%%%%%
%%%%%%%%%%%%%%%%%%%%%%%%%%%%%%%%%%%%%%%%%%%%%%
\section{Current trends} 
%%%%%%%%%%%%%%%%%%%%%%%%%%%%%%%%%%%%%%%%%%%%%%
In the search for novel approaches that could lead to the observation of more complex topology, recent theoretical works have go beyond the original proposals and extended the study to multi-terminal JJs with degenerate ground-states, which could open the door to quantum simulation of non-Abelian physics in high dimensions and holonomic quantum computing~\cite{Weisbrich2021,Xie2022}. Indeed, these devices may reveal non-Abelian Berry curvatures and higher-order Chern numbers in their non-linear response or through microwave spectroscopy.
Another twist might come from the recent realization of Floquet ABS in a graphene JJ~\cite{Park2022} which could bring the possibility of Floquet engineering into the field. 

Many researchers have looked for alternatives~\cite{Fatemi2021,Leone2008,Peyruchat2021,Herrig2022} to the multi-terminal JJ setups that might show the same  phenomena while being potentially more easy to control with current technologies. 
A promising route is the case of Weyl Josephson circuits~\cite{Fatemi2021}: arrays of coupled superconducting tunnel junctions that could be used to simulate Weyl band structures in both flux and charge space. By tailoring their design, synthetic Bloch Hamiltonians of different symmetries and dimensionalities can be engineered so as to obtain the desired topological properties. 

A completely different approach involves using gapped su\-per\-fluids of cavity exciton-polaritons in a mul\-ti-\-ter\-mi\-nal setup, as proposed in Ref. \cite{Septembre2022}. 
The advantage of these artificial systems is that topological transitions can be controlled by pump lasers with well defined phases, while the photonic component of the eigenstates, and so their topological properties, is experimentally accessible. An interesting appeal of this particular proposal is that, being exciton-polaritons a system with intrinsic gain and losses, it opens the door to non-hermitian topology. 

As we have tried to convey in this article, the study of geometrical and topological transport properties in flux networks is a research area that goes back to the early days of the quantum Hall effect~\cite{Avron1985,Avron1988,Avron1989,Avron1991}. These ideas have been brought back to the table when multi-terminal JJs were proposed as a concrete platform that could artificially mimic topological matter~\cite{Riwar2016}.
Given the promising advances in the field, one can foresee that the experimental realization of a topological flux network is not far from reach. 

\acknowledgments
We acknowledge financial support from ANPCyT (grants PICT 2016-0791, PICT 2018-1509 and PICT 2019-0371), CONICET (grant PIP 11220150100506) and from SeCyT-UNCuyo (grant 06/C603).
%%%%%%%%%%%%%%%%%%%%%%%%%%%%%%%%%%%%%%%%%%%%%%%%%%%%%%%%%%%%%%%%%%%%%%%%%%%%%%%%%%%%%%
%\bibliography{mibib}

\begin{thebibliography}{10}
\expandafter\ifx\csname url\endcsname\relax\def\url#1{\texttt{#1}}\fi

\bibitem{Klitzing1980}
\Name{Klitzing K.~v., Dorda G. \and Pepper M.} \REVIEW{Phys. Rev.
  Lett.}{45}{1980}{494}.

\bibitem{Thouless1982}
\Name{Thouless D.~J., Kohmoto M., Nightingale M.~P. \and den Nijs M.}
  \REVIEW{Phys. Rev. Lett.}{49}{1982}{405}.

\bibitem{Niu1985}
\Name{Niu Q., Thouless D.~J. \and Wu Y.-S.} \REVIEW{Phys. Rev.
  B}{31}{1985}{3372}.

\bibitem{Niu1987}
\Name{Niu Q. \and Thouless D.~J.} \REVIEW{Phys. Rev. B}{35}{1987}{2188}.

\bibitem{Avron1985}
\Name{Avron J.~E. \and Seiler R.} \REVIEW{Phys. Rev. Lett.}{54}{1985}{259}.

\bibitem{Avron2003}
\Name{Avron J.~E., Osadchy D. \and Seiler R.} \REVIEW{Physics
  Today}{56}{2003}{38}.

\bibitem{Avron1988}
\Name{Avron J.~E., Raveh A. \and Zur B.} \REVIEW{Rev. Mod.
  Phys.}{60}{1988}{873}.

\bibitem{Avron1989}
\Name{Avron J.~E. \and Cross M.~C.} \REVIEW{Phys. Rev. B}{39}{1989}{756}.

\bibitem{Avron1991}
\Name{Avron J.~E. \and Sadun L.} \REVIEW{Annals of Physics}{206}{1991}{440}.

\bibitem{Wigner1929}
\Name{Wigner E. \and von Neumann J.} \REVIEW{Phys. Z}{30}{1929}{467}.

\bibitem{Riwar2016}
\Name{Riwar R.-P., Houzet M., Meyer J.~S. \and Nazarov Y.~V.} \REVIEW{Nature
  Communications}{7}{2016}{11167}.

\bibitem{Andreev1964}
\Name{Andreev A.~F.} \REVIEW{Soviet Journal of Experimental and Theoretical
  Physics}{19}{1964}{1228}.

\bibitem{Klapwijk2004}
\Name{Klapwijk T.~M.} \REVIEW{Journal of Superconductivity}{17}{2004}{593}.

\bibitem{Nazarov2009}
\Name{Nazarov Y.~V. \and Blanter Y.~M.} \Book{Quantum {Transport}} (Cambridge
  University Press) 2009.

\bibitem{Beenakker1991}
\Name{Beenakker C. W.~J.} \REVIEW{Phys. Rev. Lett.}{67}{1991}{3836}.

\bibitem{Nielsen1981}
\Name{Nielsen H. \and Ninomiya M.} \REVIEW{Nuclear Physics B}{185}{1981}{20}.

\bibitem{Yokoyama2015}
\Name{Yokoyama T. \and Nazarov Y.~V.} \REVIEW{Phys. Rev. B}{92}{2015}{155437}.

\bibitem{Eriksson2017}
\Name{Eriksson E., Riwar R.-P., Houzet M., Meyer J.~S. \and Nazarov Y.~V.}
  \REVIEW{Phys. Rev. B}{95}{2017}{075417}.

\bibitem{Yokoyama2017}
\Name{Yokoyama T., Reutlinger J., Belzig W. \and Nazarov Y.~V.} \REVIEW{Phys.
  Rev. B}{95}{2017}{045411}.

\bibitem{Xie2018}
\Name{Xie H.-Y., Vavilov M.~G. \and Levchenko A.} \REVIEW{Phys. Rev.
  B}{97}{2018}{035443}.

\bibitem{Klees2021}
\Name{Klees R.~L., Cuevas J.~C., Belzig W. \and Rastelli G.} \REVIEW{Phys. Rev.
  B}{103}{2021}{014516}.

\bibitem{Meyer2017}
\Name{Meyer J.~S. \and Houzet M.} \REVIEW{Phys. Rev. Lett.}{119}{2017}{136807}.

\bibitem{Xie2017}
\Name{Xie H.-Y., Vavilov M.~G. \and Levchenko A.} \REVIEW{Phys. Rev.
  B}{96}{2017}{161406}.

\bibitem{PeraltaGavensky2018}
\Name{Peralta~Gavensky L., Usaj G., Feinberg D. \and Balseiro C.~A.}
  \REVIEW{Phys. Rev. B}{97}{2018}{220505}.

\bibitem{Oka2019}
\Name{Oka T. \and Kitamura S.} \REVIEW{Annual Review of Condensed Matter
  Physics}{10}{2019}{387}.

\bibitem{Houzet2019}
\Name{Houzet M. \and Meyer J.~S.} \REVIEW{Phys. Rev. B}{100}{2019}{014521}.

\bibitem{PeraltaGavensky2019}
\Name{Peralta~Gavensky L., Usaj G. \and Balseiro C.~A.} \REVIEW{Phys. Rev.
  B}{100}{2019}{014514}.

\bibitem{Xie2019}
\Name{Xie H.-Y. \and Levchenko A.} \REVIEW{Phys. Rev. B}{99}{2019}{094519}.

\bibitem{Kotetes2019}
\Name{Kotetes P., Mercaldo M.~T. \and Cuoco M.} \REVIEW{Phys. Rev.
  Lett.}{123}{2019}{126802}.

\bibitem{Sakurai2020}
\Name{Sakurai K., Mercaldo M.~T., Kobayashi S., Yamakage A., Ikegaya S., Habe
  T., Kotetes P., Cuoco M. \and Asano Y.} \REVIEW{Phys. Rev.
  B}{101}{2020}{174506}.

\bibitem{Meyer2021}
\Name{Meyer J.~S. \and Houzet M.} \REVIEW{Phys. Rev. B}{103}{2021}{174504}.

\bibitem{Ozawa2018}
\Name{Ozawa T. \and Goldman N.} \REVIEW{Phys. Rev. B}{97}{2018}{201117}.

\bibitem{Klees2020}
\Name{Klees R.~L., Rastelli G., Cuevas J.~C. \and Belzig W.} \REVIEW{Phys. Rev.
  Lett.}{124}{2020}{197002}.

\bibitem{PeraltaGavensky2022}
\Name{Peralta~Gavensky L.} \Book{Quantum transport in hybrid superconducting
  devices with topological properties.} Phd thesis UNCuyo, Instituto Balseiro
  (2022).

\bibitem{Volovik2009}
\Name{Volovik G.~E.} \Book{The Universe in a Helium Droplet} (Oxford University
  {Press}) 2009.

\bibitem{Repin2019}
\Name{Repin E.~V., Chen Y. \and Nazarov Y.~V.} \REVIEW{Phys. Rev.
  B}{99}{2019}{165414}.

\bibitem{Wang2012}
\Name{Wang Z. \and Zhang S.-C.} \REVIEW{Phys. Rev. X}{2}{2012}{031008}.

\bibitem{Wang2013}
\Name{Wang Z. \and Yan B.} \REVIEW{Journal of Physics: Condensed
  Matter}{25}{2013}{155601}.

\bibitem{Pfeffer2014}
\Name{Pfeffer A.~H., Duvauchelle J.~E., Courtois H., M\'elin R., Feinberg D.
  \and Lefloch F.} \REVIEW{Phys. Rev. B}{90}{2014}{075401}.

\bibitem{Cohen2018}
\Name{Cohen Y., Ronen Y., Kang J.-H., Heiblum M., Feinberg D., Mélin R. \and
  Shtrikman H.} \REVIEW{Proceedings of the National Academy of
  Sciences}{115}{2018}{6991}.

\bibitem{Graziano2020}
\Name{Graziano G.~V., Lee J.~S., Pendharkar M., Palmstr\o{}m C.~J. \and Pribiag
  V.~S.} \REVIEW{Phys. Rev. B}{101}{2020}{054510}.

\bibitem{Pankratova2020}
\Name{Pankratova N., Lee H., Kuzmin R., Wickramasinghe K., Mayer W., Yuan J.,
  Vavilov M.~G., Shabani J. \and Manucharyan V.~E.} \REVIEW{Phys. Rev.
  X}{10}{2020}{031051}.

\bibitem{Graziano2022}
\Name{Graziano G.~V., Gupta M., Pendharkar M., Dong J.~T., Dempsey C.~P.,
  Palmstr{\o}m C. \and Pribiag V.~S.} \REVIEW{Nature
  Communications}{13}{2022}{5933}.

\bibitem{Draelos2019}
\Name{Draelos A.~W., Wei M.-T., Seredinski A., Li H., Mehta Y., Watanabe K.,
  Taniguchi T., Borzenets I.~V., Amet F. \and Finkelstein G.} \REVIEW{Nano
  Letters}{19}{2019}{1039}.

\bibitem{Arnault2021}
\Name{Arnault E.~G., Larson T. F.~Q., Seredinski A., Zhao L., Idris S.,
  McConnell A., Watanabe K., Taniguchi T., Borzenets I., Amet F. \and
  Finkelstein G.} \REVIEW{Nano Letters}{21}{2021}{9668}.

\bibitem{Strambini2016}
\Name{Strambini E., D'Ambrosio S., Vischi F., Bergeret F.~S., Nazarov Y.~V.
  \and Giazotto F.} \REVIEW{Nature Nanotechnology}{11}{2016}{1055}.

\bibitem{Vischi2017}
\Name{Vischi F., Carrega M., Strambini E., D'Ambrosio S., Bergeret F.~S.,
  Nazarov Y.~V. \and Giazotto F.} \REVIEW{Phys. Rev. B}{95}{2017}{054504}.

\bibitem{Zgirski2011}
\Name{Zgirski M., Bretheau L., Le~Masne Q., Pothier H., Esteve D. \and Urbina
  C.} \REVIEW{Phys. Rev. Lett.}{106}{2011}{257003}.

\bibitem{Weisbrich2021}
\Name{Weisbrich H., Klees R., Rastelli G. \and Belzig W.} \REVIEW{PRX
  Quantum}{2}{2021}{010310}.

\bibitem{Xie2022}
\Name{Xie H.-Y., Hasan J. \and Levchenko A.} \REVIEW{Phys. Rev.
  B}{105}{2022}{L241404}.

\bibitem{Park2022}
\Name{Park S., Lee W., Jang S., Choi Y.-B., Park J., Jung W., Watanabe K.,
  Taniguchi T., Cho G.~Y. \and Lee G.-H.} \REVIEW{Nature}{603}{2022}{421}.

\bibitem{Fatemi2021}
\Name{Fatemi V., Akhmerov A.~R. \and Bretheau L.} \REVIEW{Phys. Rev.
  Research}{3}{2021}{013288}.

\bibitem{Leone2008}
\Name{Leone R., L\'evy L.~P. \and Lafarge P.} \REVIEW{Phys. Rev.
  Lett.}{100}{2008}{117001}.

\bibitem{Peyruchat2021}
\Name{Peyruchat L., Griesmar J., Pillet J.-D. \and Girit i. m. c.~O.}
  \REVIEW{Phys. Rev. Research}{3}{2021}{013289}.

\bibitem{Herrig2022}
\Name{Herrig T. \and Riwar R.-P.} \REVIEW{Phys. Rev.
  Research}{4}{2022}{013038}.

\bibitem{Septembre2022}
\Name{Septembre I., Meyer J.~S., Solnyshkov D.~D. \and Malpuech G.}
  \REVIEW{arXiv:2210.11088}{}{2022}{}.

\end{thebibliography}

\bibliographystyle{eplbib.bst}

\end{document}